\documentclass[12pt]{iopart}
\usepackage{graphicx}

\begin{document}

\title[Damage spreading in two dimensional geometrically frustrated lattices]{Damage spreading in two dimensional  geometrically
frustrated lattices: the triangular and
kagome anisotropic Heisenberg model}

\author{S. Bekhechi and B.W. Southern}
\address{Department of Physics and Astronomy \\University of Manitoba \\
Winnipeg Manitoba \\Canada R3T 2N2}

\date{\today}

\begin{abstract}
The  technique of damage spreading  is used to study the  phase diagram
of the easy axis anisotropic Heisenberg antiferromagnet on two geometrically 
frustrated lattices. The triangular and kagome systems are built up from triangular units
that either share  edges or corners respectively. The triangular lattice undergoes two sequential 
Kosterlitz-Thouless transitions while the kagome lattice undergoes a glassy transition. In both
cases, the phase boundaries obtained using damage spreading  are in good agreement
 with those obtained from equilibrium Monte Carlo
simulations. 
\end{abstract}

\submitto{Journal of Physics A}

\maketitle


Frustrated antiferromagnets are particularly interesting because they allow  novel kinds of low-temperature
magnetic states to develop \cite{n1a,n1b} which are quite different from those
observed in conventional magnets \cite{n1c}. When the frustration is due entirely to the geometry of the lattice, exact results
for the $S=1/2$ Ising model on the triangular \cite{n1d} and kagome \cite{n1e} lattices have shown that there is no long range order at any temperature and the
system has a macroscopic ground state degeneracy.  In the triangular case,  the
spin-spin correlation function decays with a power law \cite{n2a} whereas, in the kagome geometry, it decays
exponentially at zero temperature \cite{n2b,n2c}.
 For isotropic vector spin models (XY or Heisenberg) on a triangular lattice \cite{n2d,n3b,n3a} where the elementary triangles share edges, the frustration is partially relieved
 and a noncollinear planar ground state forms with neighbouring spins making angles of $120^o$. Finite temperature
 transitions are evident in both cases (XY and Heisenberg) but are topological in nature.
However, for the
 kagome lattice, the cornering sharing geometry leads to a disordered ground state for the Heisenberg model with the phenomenon
of order by disorder occuring in the limit $T \rightarrow 0$ with fluctuations favouring coplanar spin configurations \cite{ n3c}

In order to explore  the Heisenberg model on these geometrically frustrated 
 lattices, various types of perturbations have been applied and have shown a strong effect 
 on the ground state manifold \cite{n3d}.  We consider the following anisotropic Hamiltonian 
 
\begin{equation}
H = J\sum_{i<j} (S_{ix} S_{jx} + S_{iy} S_{jy}+ A S_{iz}
S_{jz}).
\end{equation}
where $(S_{i \alpha},\alpha=x,y,z)$ represents a classical three component
spin of unit magnitude located at each site $i$ of a  (triangular/kagome) lattice and the
exchange interactions are restricted to
nearest-neighbour pairs of sites. The parameter $A$ describes the strength of the exchange 
anisotropy. 
We restrict our attention to the case where  $A>1$ represents an easy-axis anisotropy and we measure temperature and energy
in units of $J=1$. 
The limit $A \rightarrow 1$
corresponds to the isotropic Heisenberg model  whereas the limit $A
\rightarrow \infty$ corresponds to an infinite spin Ising model.

  Monte Carlo studies have shown for the triangular lattice the existence of two distinct KT 
  types of defect-mediated phase transitions at
 finite temperature \cite{n3e,n6a} for $A>1$. By examining the limit $A \rightarrow 1$,   the transition at the Heisenberg point also appears to
 be purely topological in character \cite{n6b} but with exponentially decaying spin correlations in the low temperature phase.
Studies of the same model on the kagome lattice \cite{n6c,n6d} indicate the existence of a
 finite, but very low, temperature ferromagnetic transition  for intermediate values of $A$. Static results for the  magnetization, susceptibility and specific heat indicate an Ising-like transition but with a
 nonuniversal critical behaviour \cite{n6d} in which the values of the exponents  vary with the strength of
 the anisotropy $A$.  The individual spins do not exhibit any long ranged spatial order and resemble a glassy phase.
An analysis of the time evolution of the double time spin
 auto-correlation function
following a quench of the system from infinite temperature into a
non-equilibrium state at low temperature displays the phenomenon of aging which is a characteristic of many glassy systems.
 Additional evidence for the glassy nature
 of this low temperature state has also recently been found in a study of relationship of the spin
autocorrelation function to the  response of the system to an external magnetic
field \cite{n6e}. A violation of the fluctuation dissipation theorem is observed with properties similar to both structural glasses and spin glasses. 

\begin{figure}
\centering
\includegraphics[height=65mm,width=35mm]{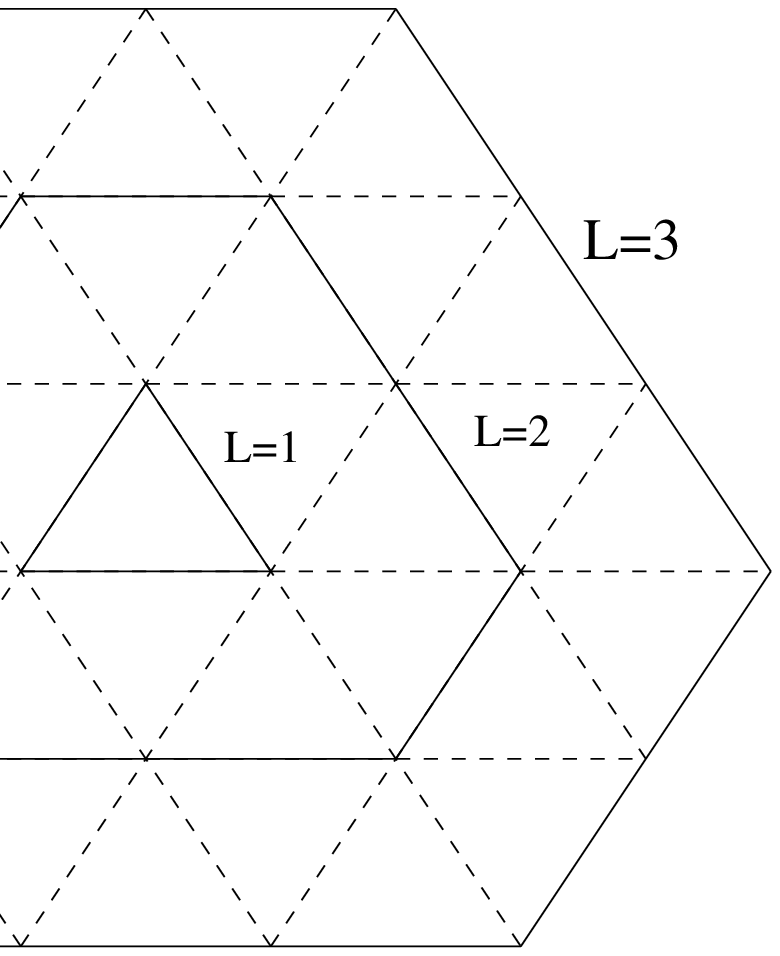}
\includegraphics[height=60mm,width=55mm]{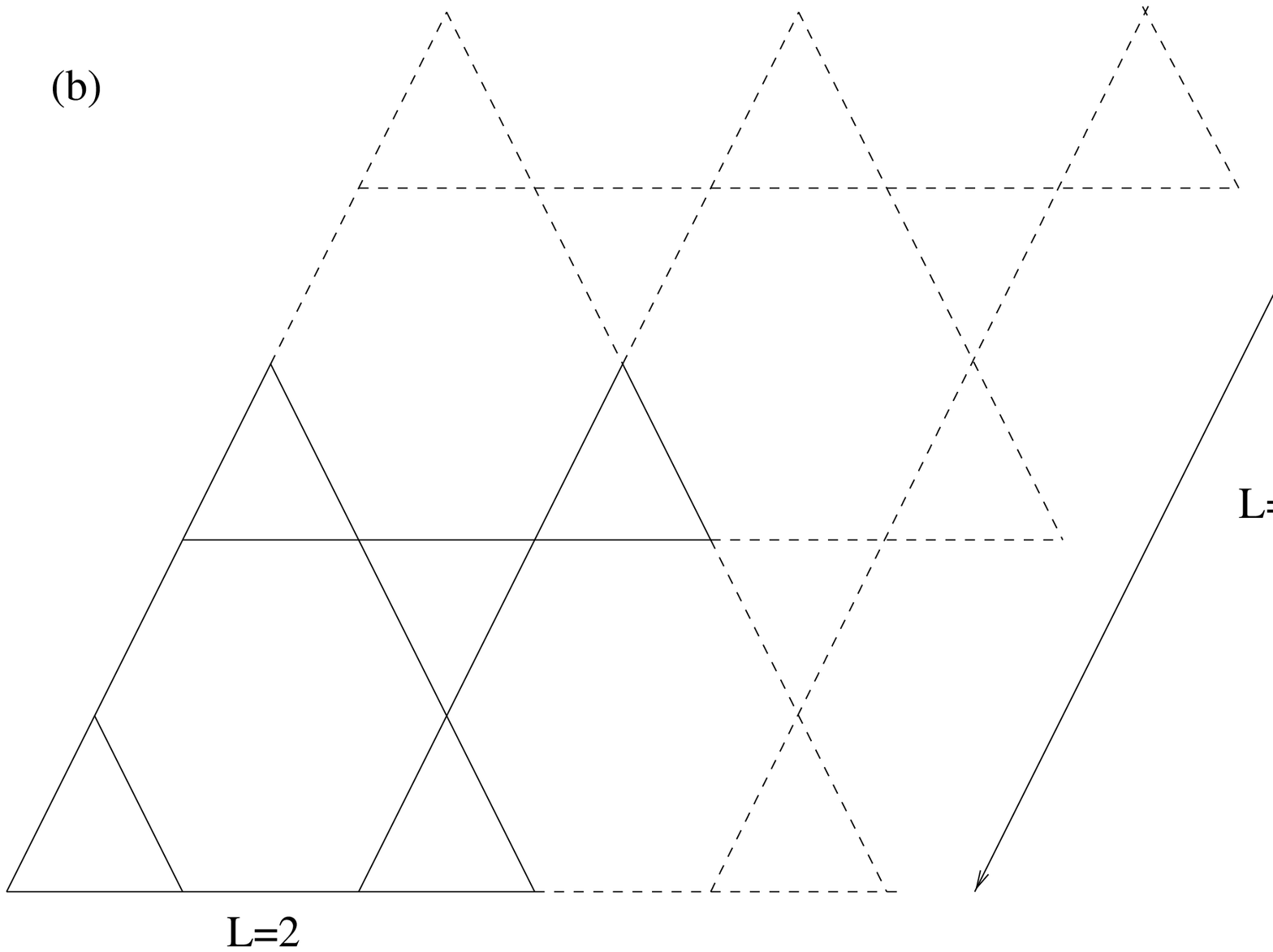}
\caption{The two geometrically frustrated lattices: (a) the triangular and (b) the kagome lattices}
\end{figure}

\begin{figure}[t]
\centering
\includegraphics[height=92mm,width=120mm]{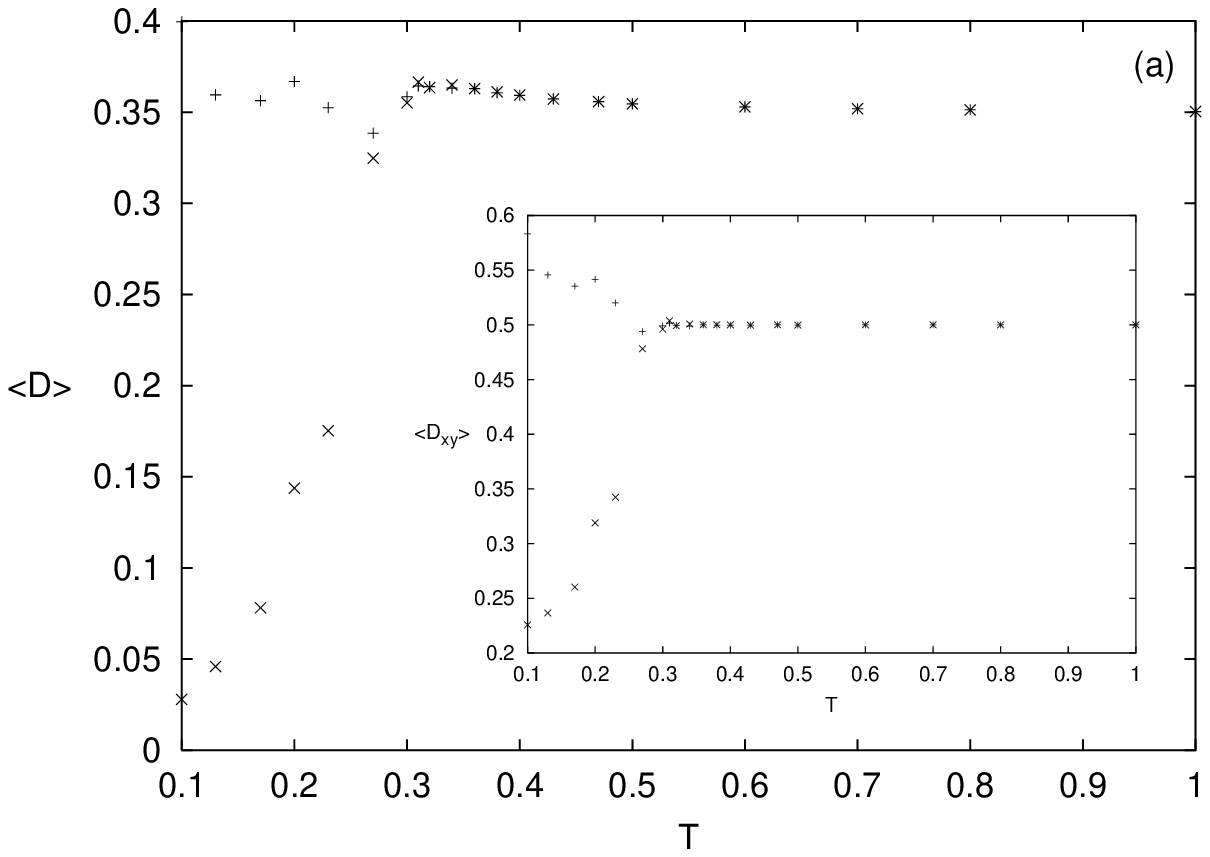}
\includegraphics[height=92mm,width=120mm]{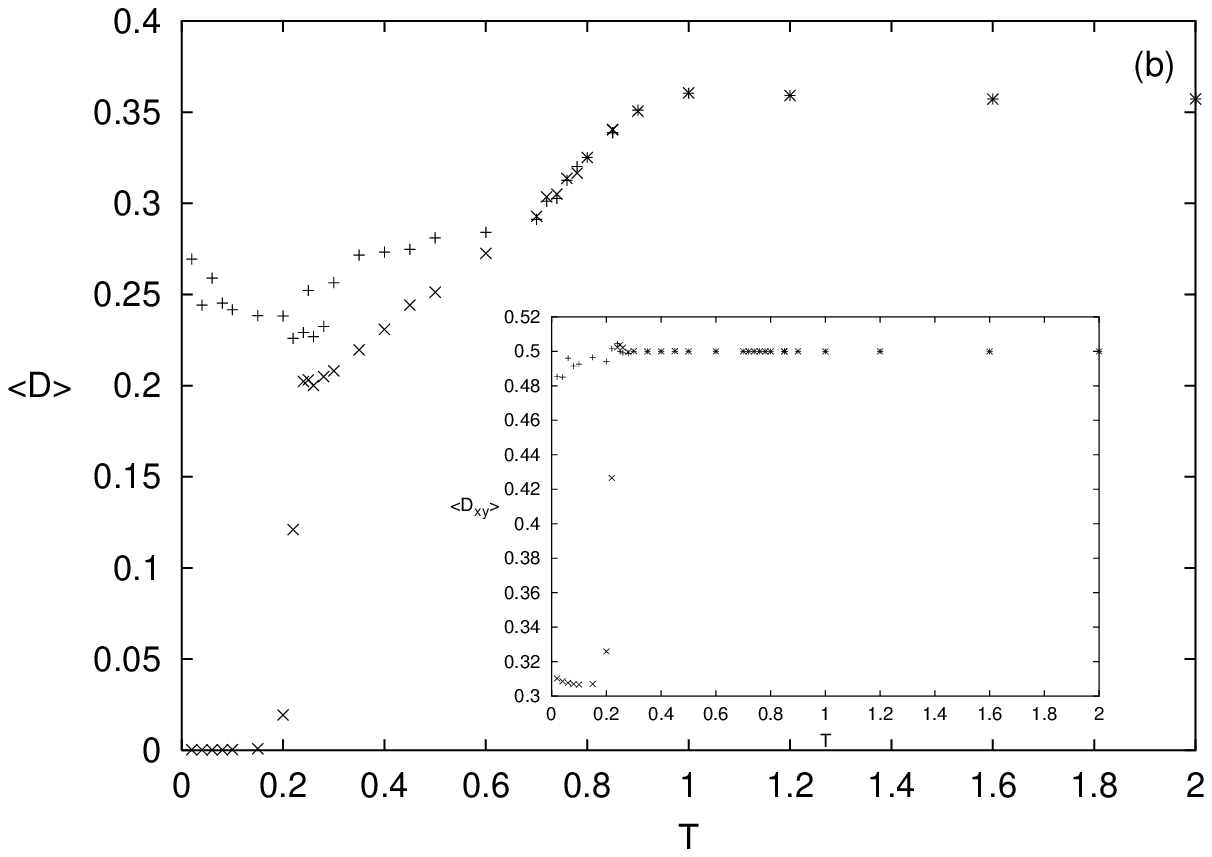}
\caption{Average  distance $<D>$ versus temperature for the triangular lattice with $L=24$ and
two differents values of the anisotropy (a) $A=1$  and (b) $A=2$ for different initial 
damage: $D(0)=1.0$(plus) and $D(0)=1/N$(crosses). The insets show
the average distance associated with the transverse components.}
\end{figure}

 In the present work we present a numerical study of  the model
 described by Hamiltonian (1) for both lattices  using a damage spreading algorithm.  In the damage spreading
 method \cite{n7a}, the damage distance between two different
 initial spin configurations is measured as they evolve with the same thermal noise.
 The distance between two configurations is defined in the following way: if
 ${\bf{S}}_{i}^{(A)}$ and ${\bf{S}}_{i}^{(B)}$ are the spins of the two different replicas,
 then the distance between them at time $t$ is defined as

  \begin{equation}
      D^{AB}(t)=\frac{1}{4N} \sum_{i=1}^{N} |{\bf{S}}_i^{(A)} -{\bf{S}}_i^{(B)}|^2
                      =\frac{1}{2N} \sum_{i=1}^{N} (1-{\bf{S}}_{i}^{(A)}. {\bf{S}}_{i}^{(B)}).
  \end{equation}
where $N$ is the total number of spins and $0 \leq D^{AB}(t) \leq 1$.  
 The procedure of the spreading of damage is as follows: for a given value of the anisotropy $A$  and temperature $T$
 we first initialize the system ${\bf{S}}_{i}^{(A)}$ in a random configuration and then let it evolve to a steady state. Then we make two additional
 copies of the system, ${\bf{S}}_{i}^{(B)}$ and ${\bf{S}}_{i}^{(C)}$. 
We introduce damage in both systems $B$ and $C$
 by inverting a fraction of the spins. The two copies $B$ and $C$ correspond to cases where
(1) a randomly chosen spin is flipped with $D^{AB}(0)=1/N$
 and (2) all the spins are flipped with $D^{AC}(0)=1$. Given these  different initial conditions we
 let all the configurations evolve in time according to the same dynamics, i.e, the same
 rule and the same random number sequence in the Monte Carlo procedure. After a
 relaxation time needed for the damaged copies to  be thermalized, we monitor the
 damage  in order to calculate the time average $<D(t)>$  of the 
 distance (2). This procedure is repeated for many different samples (initial
 configurations $A$ and damaged configurations $B$ and $C$). At high temperatures, the averages $<D^{AB}>$ and $<D^{AC}>$ are
equal but at low temperatures they approach different values. We interpret this result as indicating that the free energy landscape
has a single valley at high temperature and that a transition to a more complicated multi-valley landscape occurs as the
temperature is lowered.

Figure 1 shows the geometry of the two lattices  for different sizes $L$.  In both cases the number of
sites $N=3L^2$ and periodic boundary conditions are applied. 
Our simulations have been performed for  sizes ranging from $L=12$ to $L=60$, but
 all of the results are presented with $L=24$ where finite size effects are sufficiently small and
  a larger number of Monte Carlo sweeps (MCS) are possible.
 For each temperature we calculate the time average of the damage  using 10000 MCS and
 typically 200 samples were used to compute the averages.
 
\begin{figure*}[hbp]
\begin{minipage}{75mm}
\centering
\includegraphics[ height=65mm,width=75mm ]{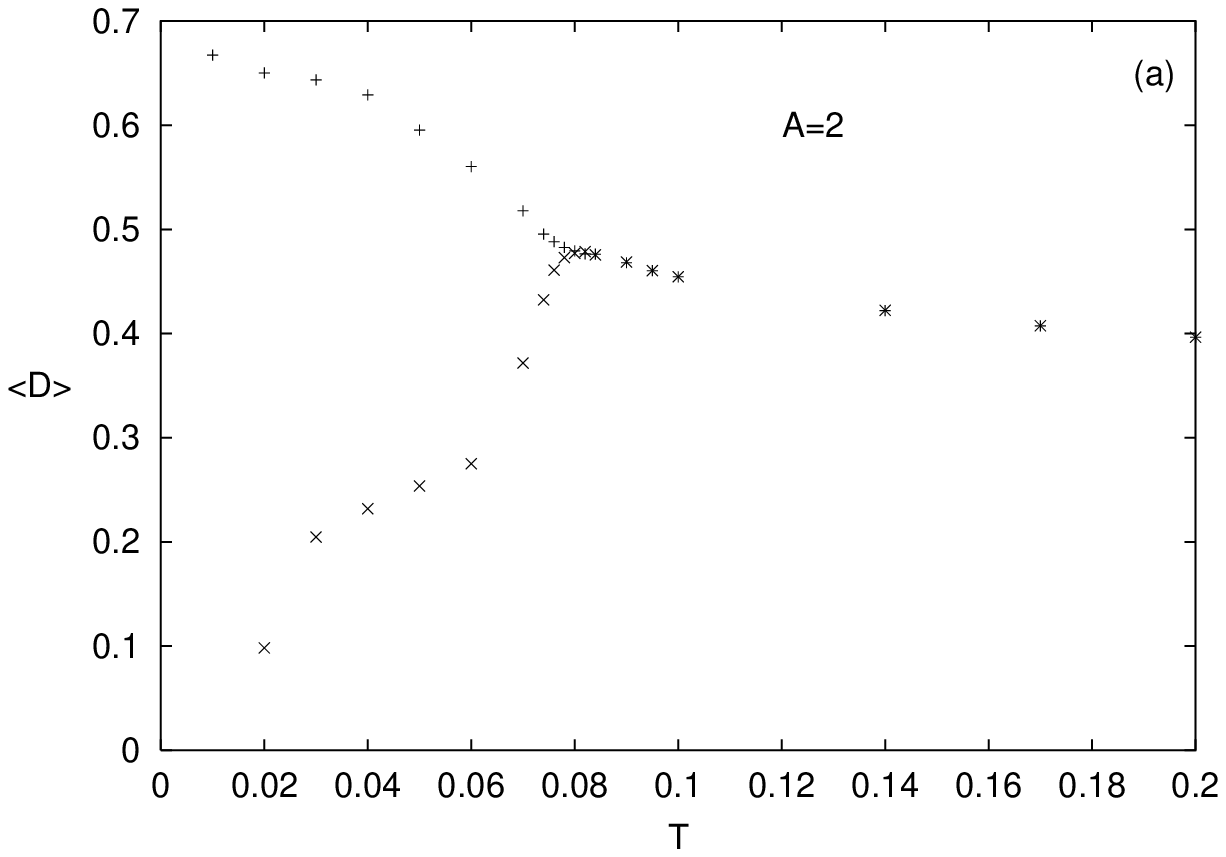}
\end{minipage}
\hspace{10pt}
\begin{minipage}{75mm}
\center{\includegraphics[height=65mm,width=75mm]{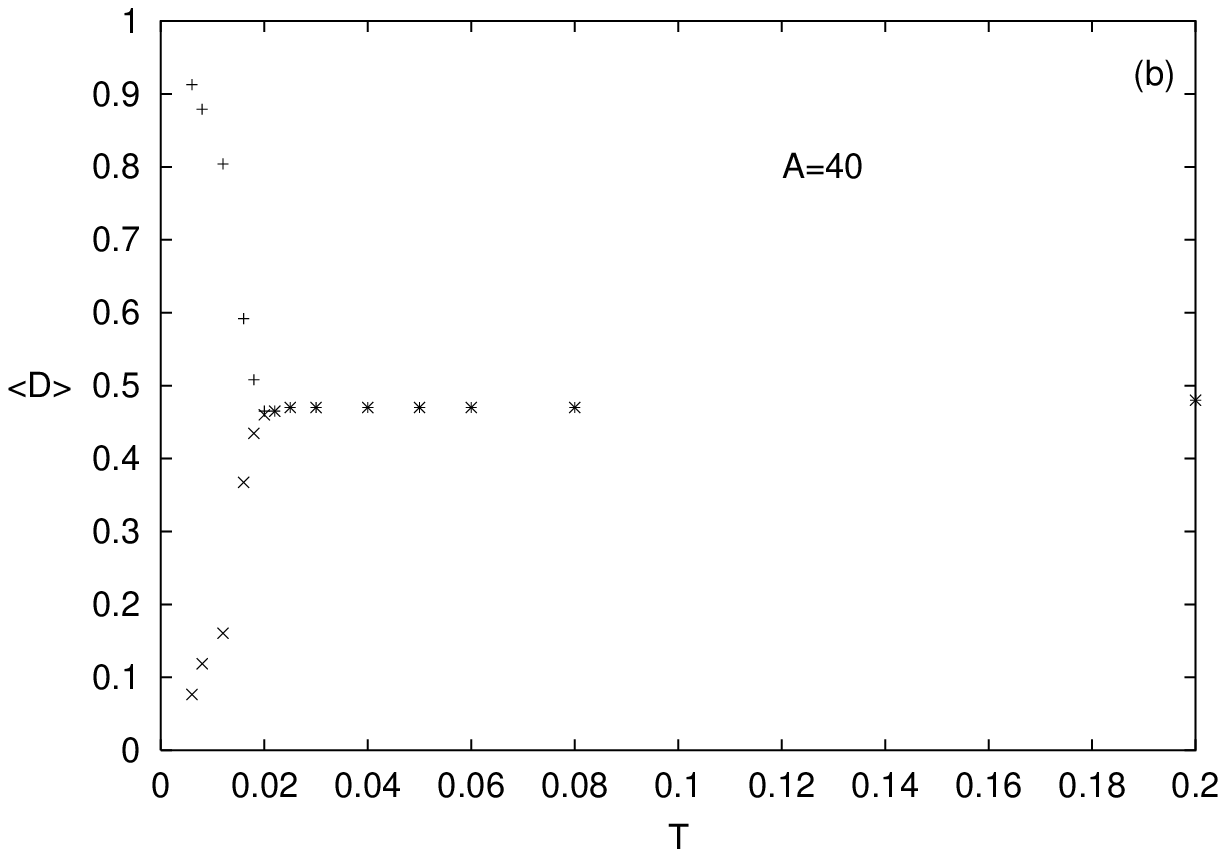}}
\end{minipage}
\caption{Averaged  distance $<D>$ versus temperature for the kagome 
lattice with $L=24$ and
two different values of the anisotropy (a) $A=2$ and (b) $A=40$  for
 the  initial damage $D(0)=1.0$(plus) and $D(0)=1/N$(crosses).
}
\end{figure*}

 In the case of the triangular lattice,  recent Monte
 Carlo simulations \cite{n6b} have shown the existence of two finite temperature KT transitions. 
The upper transition  $T_{c2}$ corresponds to the onset of power law correlations for the spin components
parallel to the easy axis $A>1$ and the lower transition  $T_{c1}$ corresponds to the onset of power law
correlations for the perpendicular components. In both cases there is a corresponding spin stiffness
coefficient which vanishes. In order to identify both transitions,  we also compute the following
measure of the damage distance associated 
with the transverse (xy) components of the spins in the two replicas

\begin{equation}
D_{xy}^{AB}= \frac{1}{2N} \sum_{i=1}^{N} (1-{{S_{ix}^{(A)}}} {{S_{ix}^{(B)}}}-{S_{iy}^{(A)}} {{S_{iy}^{(B)}}}).
\end{equation}

 As shown in  figure 2(a) for the isotropic Heisenberg case $(A=1)$ both distances
  are independent of the initial damage at high temperature but become dependent on it below the  same temperature $T_c \sim 0.31$
which agrees quite well with previous estimates of the transition temperature using equilibrium methods \cite{n2d,n3b}. However, for values of $A>1$, three
 different temperature regions are observed as shown  in figure 2(b) for the case $A=2$ :
 (i) $T<T_{c1}$, both distances depend on the initial damage,  
 (ii) $T_{c1}<T<T_{c2}$, $<D_{xy}>$ is independent of the initial damage whereas
 $<D>$  still depends on the initial damage and  
 (iii)  $T>T_{c2}$, both distances become independent on the initial damage.
The damage spreading approach is able to identify both transitions which, in this case, correspond to defect unbinding transitions
and it also predicts a finite temperature transition in the Heisenberg limit.

\begin{figure*}[hbp]
\begin{minipage}{75mm}
\centering
\includegraphics[height=65mm,width=75mm,]{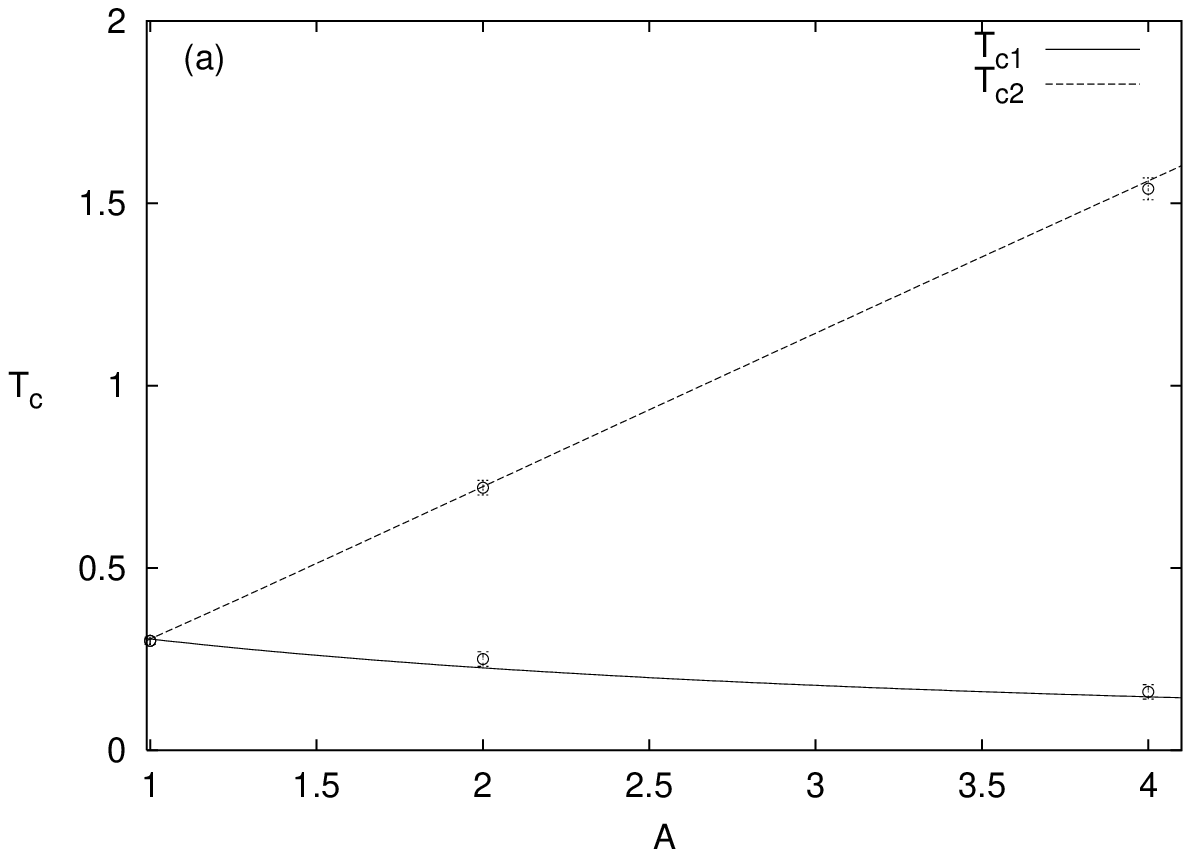}
\end{minipage}
\hspace{10pt}
\begin{minipage}{70mm}
\center{\includegraphics[height=65mm,width=75mm]{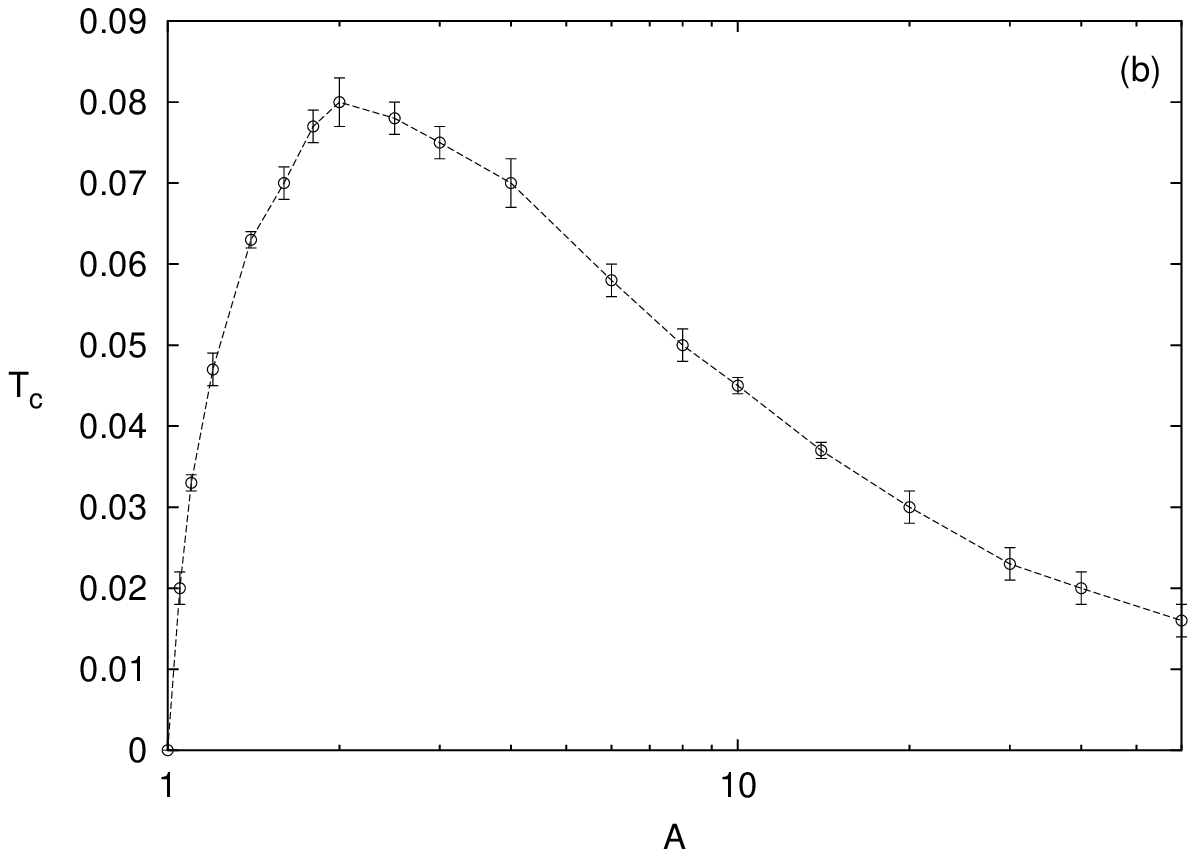}}
\end{minipage}
\caption{Phase diagrams  obtained from the damage spreading method for (a) the triangular and (b) 
the kagome lattices respectively. The two lines in (a) correspond to the equations 
$T_{c1}= T_c \frac {4(2A+1)}{3(1+A)^2}$ and $T_{c2}= T_c \frac {4A^2(A+2)}{3(1+A)^2}$ with 
$T_c=(0.305\pm 0.005)$ \cite{n6a}. 
}

 \end{figure*}

 Our results for the Kagome lattice are shown in figure 3. In this case there are only two phases: a low
 temperature phase $T<T_c$ where the value of $<D>$ is dependent on the initial value of $D$ and
 a high temperature phase $T>T_c$ where  $<D>$ is independent of the initial
 damage. The transverse distance $D_{xy}$ does not indicate any transition at all. The value of $T_c$ depends on the value of the anisotropy $A$ and approaches zero in both
the $A \rightarrow 1$ and $A \rightarrow \infty$ limits. In figure 4, we show the phase diagrams obtained using this
damage spreading approach for both lattices. Both phase
 diagrams are in remarkably good agreement with those obtained from equilibrium
 studies \cite{n3e,n6a,n6c,n6e}  within errorbars.

 In summary, we have shown that the technique of damage spreading can be 
 applied successfully to models where analytical solutions are lacking and where
 the low temperature phase displays novel  behaviour associated with  geometrical
frustration. We can clearly identify the damage
 transitions with those obtained previously using conventional Monte Carlo simulations.
Both defect-mediated and glass-like transitions are detected by the method.  Further work is
needed to extract more detailed information about the critical properties using this approach.\\
 
 {\bf Acknowledgments}

 The authors would like to acknowledge  support from the Natural Sciences and
 Research Council of Canada and  the High
 Performance Computing Facility at the University of Manitoba.

\end{document}